\documentclass[fleqn,10pt]{wlscirep}

\usepackage{algorithm}
\usepackage{algpseudocode}
\usepackage{amsmath}
\usepackage{amsfonts}
\usepackage{array}
\usepackage{caption}
\usepackage{comment}
\usepackage{graphicx}
\usepackage{indentfirst} 
\usepackage{lineno,hyperref}
\usepackage{mathrsfs}
\usepackage{multirow}
\usepackage{rotating}
\usepackage{siunitx}
\usepackage{soul}
\usepackage{subcaption}
\usepackage{tabularx}
\usepackage[normalem]{ulem}
\usepackage{url}
\usepackage{xcolor}

\newcounter{barmanEnum}
\newcounter{figsInText} 
\newcommand{\no}{\noindent}
\newcommand{\bsym}{\boldsymbol}

\newcommand{\bigstretch}{1.25}
\newcommand{\p}{\mathcal{P}^2}

\begin{document}
	\title{Study of the convergence of the Meshless Lattice Boltzmann Method in Taylor-Green and annular channel flows}
	
	\author[1,*]{Dawid Strzelczyk}
	\author[1]{Maciej Matyka}
	\affil[1]{Institute of Theoretical Physics,\newline
		Faculty of Physics and Astronomy,\newline University of Wroc\l aw, \newline pl. M. Borna 9, 50-204, Wroc\l aw, Poland}
	
	\affil[*]{dawid.strzelczyk@uwr.edu.pl}
	
	\begin{abstract}
		The Meshless Lattice Boltzmann Method (MLBM) is a numerical tool that relieves the standard Lattice Boltzmann Method (LBM) from regular lattices and, at the same time, decouples space and velocity discretizations. In this study, we investigate the numerical convergence of MLBM in two benchmark tests: the Taylor-Green vortex and annular (bent) channel flow. We compare our MLBM results to LBM and to the analytical solution of the Navier-Stokes equation. We investigate the method's convergence in terms of the discretization parameter, the interpolation order, and the LBM streaming distance refinement. We observe that MLBM outperforms LBM in terms of the error value for the same number of nodes discretizing the domain. We find that LBM errors at a given streaming distance $\delta x$ and timestep length $\delta t$ are the asymptotic lower bounds of MLBM errors with the same streaming distance and timestep length. Finally, we suggest an expression for the MLBM error that consists of the LBM error and other terms related to the semi-Lagrangian nature of the discussed method itself.
	\end{abstract}
	
	\keywords{Lattice Boltzmann Method, meshless methods,  radial basis functions, convergence analysis}
	
	\maketitle
	
	
	\section{Introduction}\label{sec:introduction}

	A challenging task preceding most computational fluid dynamics calculations is discretizing the domain. It often consists of complex geometries, especially in flows through porous media \cite{Andrade1999}, biological structures \cite{Jin2016}, past cars \cite{Fu2018} and marine vehicles \cite{Jasak2019} which makes the use of irregular discretizations inevitable. In such cases, standard approaches, such as the Finite Volume Method, need to perform intensive computations at the level of domain discretization, not only to place nodes in the domain but also to connect them in such a way that they tessellate the space properly \cite{Baker2005}. Thus, additional memory is required to store information about the connectivity between nodes. Moreover, local, adaptive refinement of the discretization, which aims to increase the solution accuracy only where necessary, implies costly mesh manipulations even during the simulation runtime. A remedy for these issues is to use meshless methods in which domains are discretized with clouds of non-connected points \cite{Liu2005}. They allow for more freedom in the node placement in the domain, a less costly local refinement procedure due to the lack of node connectivity information, and feasible formulation of high-order approximations. However, this comes at the cost of, for example, conservation laws not being imposed explicitly on the solved equations. One of the meshless methods is the radial basis functions (RBF) method \cite{Bayona2015,Slak2019,Fornberg2015} where interpolation and derivatives approximation are performed in radial functions basis.
	
	One does not need to solve the Navier-Stokes equation directly to obtain quantities such as fluid velocity and pressure. The desired quantities can be calculated by solving Boltzmann’s transport equation (BTE). Lattice Boltzmann Method (LBM)\cite{Succi2018} is a numerical tool for solving BTE that has in the last decades become a popular method of choice for simulating transport phenomena. It is advantageous compared to the direct Navier-Stokes solution methods in terms of stability, ease of parallelization, and simplicity of implementation. Applications range from porous media flows \cite{Matyka2021,Koponen1997,Andrade1999}, multiphase flows~\cite{Paradis2013,Gu2021} including cavitating flows~\cite{Falcucci2013} to the flow of semiclassical fluids \cite{Coelho2018} and relativistic hydrodynamics \cite{Mendoza2010}. In its standard formulation, the LBM operates on regular square (in 2D) discretization (lattices) of the space, which is related to the discretization of particle velocities. There are numerous works where the standard algorithm is extended to use unstructured discretizations to allow for a better representation of the underlying geometry in the model and provides more freedom for the placement of the nodes inside the domain. For example in \cite{He1997} a cylindrical mesh with second order interpolation is used to simulate flow around a cylinder, whereas in \cite{He1996} rectangular meshes of various densities along with linear and quadratic interpolation are used to model the flow in a channel behind a sudden contraction. A method that utilizes Lagrange polynomials of various orders in periodic flows was investigated in \cite{Kramer2017}. The approaches put forward in \cite{Misztal2015} and \cite{Bardow2006} use finite elements to solve the weak forms of the discrete Boltzmann equation. A more general discussion on solving BTE with finite elements on unstructured discretizations is presented in \cite{Lee2003}.
	
	The authors of \cite{Lin2019} develop an MLBM algorithm capable of operating on clouds of scattered, non-connected points instead of grids. This method uses radial basis function interpolation to solve the streaming step in a semi-Lagrangian way. Although the original method seems promising and proves the numerical accuracy of the obtained results of the flow, its convergence toward the analytical solution was neither tested nor discussed. Thus, in this work, we aim to address the question of the spatial convergence of MLBM presented in \cite{Lin2019}. 
	In contrast to the original work, where regular or O-type grids were used, we perform a systematic study using node sets more suited for meshless methods. For the benchmark, we choose two distinct flows: the Taylor-Green vortex and a flow in an annular channel. For both, we use irregular spatial discretizations generated by the algorithm from \cite{Slak2019a} based on Poisson disk sampling. Our results show, that although LBM errors are in principle smaller, there are regions in which MLBM outperforms it and may be a good alternative, especially in systems with irregular boundaries.
	
	\section{Methods}\label{sec:methods}
	
	\subsection{Lattice Boltzmann Method}\label{subsec:lbm}
	
	Lattice Boltzmann Method \cite{Succi2018, Kruger2017} solves equations governing the evolution of discrete velocities distribution function:
	\begin{equation}\label{eq:LBM_streaming}
		f_k(t+1,\boldsymbol{x}) = f_k^\text{post}(t,\boldsymbol{x}+\boldsymbol{e}_{k'})
	\end{equation}
	where $f_k$ is the distribution function associated with the $k$-th streaming vector $\bsym{e}_k$, $k'$ denotes the direction opposite to $k$ ($\boldsymbol{e}_k=-\boldsymbol{e}_{k'}$) and the superscript 'post' denotes the post-collision distribution function. According to this notation, $\boldsymbol{x}+\boldsymbol{e}_{k'}$ denotes the neighbor of lattice node $\boldsymbol{x}$ lying one lattice site in the upstream direction $\boldsymbol{e}_{k}$ from this node. Note that Eq. \eqref{eq:LBM_streaming} is written in non-dimensionalized form with the timestep length equal to 1. 
	We use D2Q9 BGK model with 9 streaming directions ($k=0,1,...,8$):
	\begin{equation}
		\boldsymbol{e}_k \in \left(
		(0,0),
		(1,0),
		(0,1),
		(-1,0),
		(0,-1),
		(1,1),
		(-1,1),
		(-1,-1),
		(1,-1)
		\right)
	\end{equation}
	which imposes the use of a square lattice. Refer to Fig.~\ref{fig:lbm_discretization} for a graphical representation of the LBM lattice and streaming directions. The mentioned model uses $f^\text{post}_k$ in the form suggested by Bhatnagar, Gross and Krook (BGK)~\cite{Bhatnagar1954}:
	\begin{equation}\label{eq:LBM_collision}
		f^\text{post}_k = f_k(t,\boldsymbol{x}) - \frac{1}{\tau}\left[f_k(t,\boldsymbol{x}) - f^\text{eq}_k(t,\boldsymbol{x})\right]
	\end{equation}
	where $\tau$, the so-called \textit{relaxation time}, determines the characteristic timescale of the particle populations reaching local equilibrium. The equilibrium distributions $f_k^\text{eq}$ are expressed as:
	\begin{equation}\label{eq:feq}
		f^\text{eq}_k(t,\boldsymbol{x}) = \omega_k\rho\left[1+\frac{\boldsymbol{c}_k\cdot \boldsymbol{u}}{c_s^2} + \frac{(\boldsymbol{c}_k\cdot \boldsymbol{u})^2}{2c_s^4} - \frac{\boldsymbol{u}^2}{2c_s^2}\right]
	\end{equation}
	where $c_s=1/\sqrt{3}$ denotes the lattice speed of sound. $\omega_k$ is the weight specific to the $k$th streaming direction:
	\begin{equation}
		\omega = \left(\frac{4}{9},\frac{1}{9},\frac{1}{9},\frac{1}{9},\frac{1}{9},\frac{1}{36},\frac{1}{36},\frac{1}{36},\frac{1}{36}\right).
	\end{equation}
	The nondimensionalized kinematic viscosity of the modeled fluid can be calculated as $\nu_{lb}=(\tau-0.5)/c_s^2$. Macroscopic density and velocity, $\rho\!=\!\rho(t,\boldsymbol{x})$ and $\boldsymbol{u}\!=\!\boldsymbol{u}(t,\boldsymbol{x})$ respectively, at time $t$ and point $\boldsymbol{x}$, are obtained from discrete populations:
	\begin{equation}\label{eq:macro_var}
		\renewcommand{\arraystretch}{2.5}
		\begin{array}{l}
			\rho = \sum\limits_{k=1}^q f_k \\
			\bsym{u} = \frac{1}{\rho}\sum\limits_{k=1}^q f_k \bsym{e}_k
		\end{array}
	\end{equation}
	In the numerical implementations, at each time step, Eq. ~\eqref{eq:LBM_collision} is first calculated to obtain the values of the post-collision distribution function (\textit{collision step}). Then, ~\eqref{eq:LBM_streaming} advects post-collision distributions to neighboring nodes (\textit{streaming step}). Because lattice nodes $\bsym{x}$ coincide with the departure/arrival nodes of the streaming step, transport is purely Lagrangian and amounts to an index shift in the distribution function array.
	
	\subsection{Radial Basis Functions interpolation}\label{subsec:rbf}
	
	We use meshless interpolation in radial functions basis to perform the semi-Lagrangian streaming step of the LBM. Consider a set of $N$ points in $d$-dimensional Euclidean space $\boldsymbol{x}_i\!\in\!\mathbb{R}^d,\,\,\,i=1,2,\dots,N$ forming a set $X=\{\bsym{x}_1,\bsym{x}_2,\dots,\bsym{x}_N \}$. Each point can be assigned a set $X^L \subset X$, $L=1,...,N$ of its closest neighbors, including the node itself. We denote $N_L$ members of $X^L$ as $\bsym{x}^L_i$, $i=1,...,N_L$. We assume that in every set $X^L$ the first node $\bsym{x}^L_1 \equiv \bsym{x}_i$ and we call this node the {\itshape stencil center}. We also assume that, at every point $\bsym{x} \in X$ we know the value $f_i \equiv f(\bsym{x}_i)$ of the interpolated function $f$. In our case, these are the nine distribution functions introduced in the previous section. In each of $X^L$ sets one can construct local RBF interpolants $F_L$:
	\begin{equation}\label{eq:interpolant_value_alg}
		F_L(\bsym{x}) = \sum\limits_{i=1}^{N_L} \gamma^L_i \phi^L_i(\bsym{x}) + \sum\limits_{j=1}^{N^P_L} \pi^L_j p^L_j(\bsym{x}) \approx f(\bsym{x})
	\end{equation}
	where $\phi_i^L(\bsym{x}) \equiv \phi_i(|\bsym{x}-\bsym{x}^L_i|) \equiv \phi_i(r)$ and $p^L_j(\bsym{x})$ are radial functions and polynomials forming the RBF and polynomial subsets of the interpolation basis, respectively. $\gamma^L_i$ and $\pi^L_i$ are coefficients of the linear combination of the interpolation basis elements. It can be written in vector form as:
	\begin{equation}\label{eq:interpolant_value}
		F_L(\bsym{x}) = \left[\bsym{\gamma}_L^T, \bsym{\pi}_L^T\right] \cdot \bsym{\phi}_L \approx f(\bsym{x})
	\end{equation}
	where coefficient vectors $\bsym{\gamma}_L$ and $\bsym{\pi}_L$ are given as:
	\begin{equation}
		\bsym{\gamma}_L
		=
		\left\{
		\def\arraystretch{\bigstretch}
		\begin{array}{c}
			\gamma^L_1\\
			\gamma^L_2\\
			\vdots\\
			\gamma^L_{N_L}\\
		\end{array}
		\right\}, \quad
		\bsym{\pi}_L
		=
		\left\{
		\def\arraystretch{\bigstretch}
		\begin{array}{c}
			\pi^L_1\\
			\pi^L_2\\
			\vdots\\
			\pi^L_{N^P_L}\\
		\end{array}
		\right\}
	\end{equation}
	and $\bsym{\phi}_L$ is the vector of $X^L$'s basis functions values at point $\bsym{x}$:
	\begin{equation}
		\bsym{\phi}_L(\bsym{x})
		=
		\left[
		\phi^L_1(\bsym{x}),
		...
		\phi^L_{N_L}(\bsym{x}),
		p^L_1(\bsym{x}),
		...
		p^L_{N^P_L}(\bsym{x})
		\right]^T
	\end{equation}
	Vectors $\bsym{\gamma}_L$ and $\bsym{\pi}_L$ are determined in each set $X^L$ by the following set of {\itshape collocation equations}:
	\begin{equation}\label{eq:rbf_interp_set_augmented}
		\bsym{\Phi}_L \cdot 
		\left\{
		\begin{array}{c}
			\boldsymbol{\gamma}_L \\
			\boldsymbol{\pi}_L
		\end{array}
		\right\} =
		\begin{bmatrix}
			\boldsymbol{R}_L & \boldsymbol{P}_L \\
			\boldsymbol{P}_L^T & \boldsymbol{0}
		\end{bmatrix}
		\cdot
		\left\{
		\begin{array}{c}
			\boldsymbol{\gamma}_L \\
			\boldsymbol{\pi}_L
		\end{array}
		\right\}
		=
		\left\{
		\begin{array}{c}
			\boldsymbol{f}_L \\
			\boldsymbol{0}
		\end{array}
		\right\}
	\end{equation}
	where $\bsym{0}$ is either a $N^P_L \times N^P_L$ matrix of zeros or $N^P_L$-element vector of zeros and $\bsym{f}_L$ is an array of the interpolated function values at $X^L$ members:
	\begin{equation}
		\bsym{f}_L = \left[f(\bsym{x}^L_1),...,f(\bsym{x}^L_{N_L})\right]^T.
	\end{equation}
	The RBF interpolation matrix $\bsym{R}^L$ and the polynomial interpolation matrix $\bsym{P}_L$ are given as
	\begin{equation}
		\begin{array}{l}	
			\bsym{R}_L
			=
			\def\arraystretch{\bigstretch}
			\begin{bmatrix}
				\phi_1^L(\bsym{x}^L_1) & \phi_2^L(\bsym{x}_1^L) & \cdots & \phi_{N_L}^L(\bsym{x}_1^L) \\
				\phi_1^L(\bsym{x}_2^L) & \phi_2^L(\bsym{x}_2^L) & \cdots & \phi_{N_L}({\bsym{x}}_2^L) \\
				\vdots & \vdots & \ddots & \vdots \\
				\phi_1^L(\bsym{x}^L_{N_L}) & \phi_2^L(\bsym{x}^L_{N_L}) & \cdots & \phi_{N_L}^L(\bsym{x}^L_{N_L}) \\
			\end{bmatrix} \\ \\
			
			\bsym{P}_L
			=
			\def\arraystretch{\bigstretch}
			\begin{bmatrix}
				p_1^L(\bsym{x}^L_1) & p_2^L(\bsym{x}_1^L) & \cdots & p_{N^P_L}^L(\bsym{x}_1^L) \\
				p_1^L(\bsym{x}_2^L) & p_2^L(\bsym{x}_2^L) & \cdots & p_{N^P_L}(\bsym{x}_2^L) \\
				\vdots & \vdots & \ddots & \vdots \\
				p_1^L(\bsym{x}^L_{N_L}) & p_2^L(\bsym{x}^L_{N_L}) & \cdots & p_{N^P_L}^L(\bsym{x}^L_{N_L}) \\
			\end{bmatrix}
		\end{array}
	\end{equation}
	Transforming Eq.~\eqref{eq:rbf_interp_set_augmented}, the coefficient vectors can be written as:
	\begin{equation}
		\left\{
		\begin{array}{c}
			\boldsymbol{\gamma}_L \\
			\boldsymbol{p}_L
		\end{array}
		\right\}
		=
		\bsym{\Phi}^{-1}
		\cdot
		\left\{
		\begin{array}{c}
			\boldsymbol{f}_L \\
			\boldsymbol{0}
		\end{array}
		\right\}
	\end{equation}
	with the use of which one can rewrite Eq.~\eqref{eq:interpolant_value} as:
	\begin{equation}
		F_L(\bsym{x}) = \left(
		\bsym{\Phi}^{-1}
		\cdot
		\left\{
		\begin{array}{c}
			\boldsymbol{f}_L \\
			\boldsymbol{0}
		\end{array}
		\right\}
		\right)
		\cdot
		\bsym{\phi}_L(\bsym{x}) \approx f(\bsym{x})
	\end{equation}
	and by rearranging terms:
	\begin{equation}\label{eq:shape_vector}
		\hspace{-.5cm}
		F_L(\bsym{x}) = \left(
		\bsym{\Phi}^{-1}
		\cdot
		\bsym{\phi}_L(\bsym{x})
		\right)
		\cdot
		\left\{
		\begin{array}{c}
			\boldsymbol{f}_L \\
			\boldsymbol{0}
		\end{array}
		\right\}
		=
		\bsym{w}_L(\bsym{x})
		\cdot
		\left\{
		\begin{array}{c}
			\boldsymbol{f}_L \\
			\boldsymbol{0}
		\end{array}
		\right\}
		\approx f(\bsym{x})
	\end{equation}
	In this manner, one defines the value of interpolant $F_L$ at an arbitrary point $\bsym{x}$ in terms of the interpolated function values $\bsym{f}_L$ and the \textit{shape vector} $\bsym{w}_L(\bsym{x})$.
	
	\subsection{Meshless LBM with RBF interpolation}\label{subsec:olbm}
	
\begin{figure}[h!]
		\centering
		\includegraphics[width=.85\linewidth]{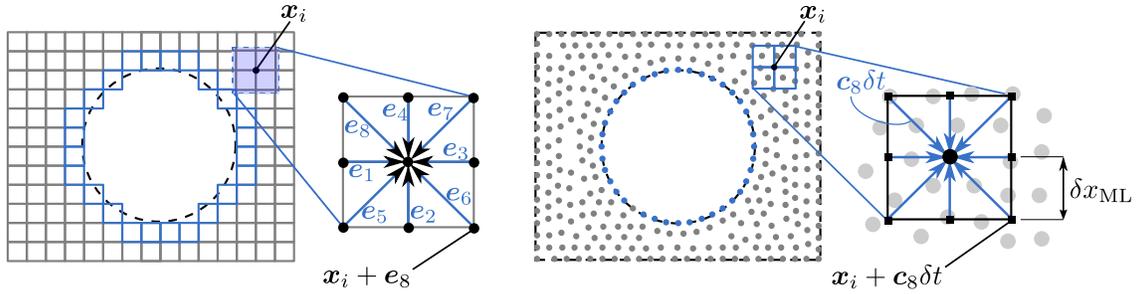}
		\caption{{\itshape Left}: a regular LBM grid discretizing a circle (dashed line) with a zoom at a local neighborhood of a discretization node. Note that the boundary is discretized in a stair-shaped manner (blue squares). {\itshape Right}: a meshless discretization of the same geometry results in nodes lying exactly on the boundary. Local neighborhoods are build for each node similar to the lattice approach but with the streaming directions $\bsym{e}_k$ replaced by streaming velocities $\bsym{c}_k$.}
		\label{fig:lbm_discretization}
	\end{figure}

	We use the D2Q9 BGK LBM model and solve the streaming step in a semi-Lagrangian way using RBF interpolation. Because departure nodes ($\boldsymbol{x}+\boldsymbol{e}_{k'}$ in Eq. \eqref{eq:LBM_streaming}) need no longer to coincide with the lattice points we find it more convenient to introduce streaming length $\delta x_\text{ML}$ and timestep length $\delta t$, both in physical units, to describe the model. In this manner lattice directions $\bsym{e}_k$ are substituted with \textit{lattice velocities} $\bsym{c}_k=\bsym{e}_k\delta x_\text{ML} / \delta t$. To relate meshless and standard LBM setups to one another, one can introduce the standard LBM discretization parameter $\delta x\!=\!L_0/N_L$ where $L_0$ is some physical reference length and $N_L$ is the number of lattice nodes discretizing it and compare $\delta x_\text{ML}$ with $\delta x$. It is also necessary to distinguish between \textit{Eulerian nodes}, where the values of distribution functions are stored and where collisions take place, and \textit{Lagrangian nodes}, which serve only as departure nodes for distribution functions during streaming (see the right subplot of Fig. \ref{fig:lbm_discretization}).
	
\begin{figure}[h!]
		\centering
		\includegraphics[width=.7\linewidth]{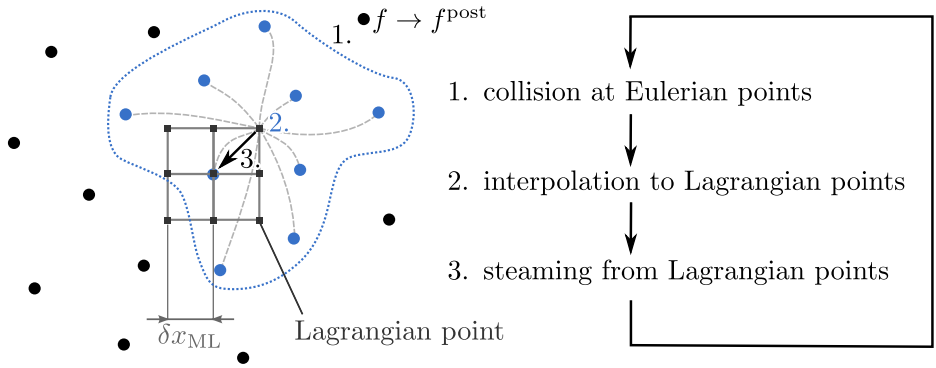}
		\caption{The procedure performed at a single timestep of the meshless LBM algorithm. Circles represent Eulerian points, squares denote Lagrangian points. A dashed loop encloses the stencil of a Lagrangian node which consists of the 9 closest neighbors of the Lagrangian node (blue circles). The interpolated distribution function is streamed to the Eulerian point lying at the center of the presented square lattice.}
		\label{fig:olbm_basics}
	\end{figure}

	In the semi-Lagrangian approach, the collision step is performed, due to its local nature, in the same way as in the standard LBM - in each Eulerian point. Streaming step, solved in a semi-Lagrangian way consists of interpolating the values of distribution functions from Eulerian to Lagrangian nodes and then advecting the interpolated distributions to the target Eulerian nodes. Refer to Figure \ref{fig:olbm_basics} for a graphical interpretation of the above three steps. Giving up the Lagrangian nature of the streaming step decouples velocity and space discretizations since Lagrangian nodes positions $\bsym{x} + \bsym{c}_k\delta t = \bsym{x} + \delta \bsym{x}_{k'}$ need not to coincide with any of Eulerian nodes.
	
	To perform interpolation we find the closest Eulerian neighbor of each Lagrangian point and determine the members of its stencil. Then, the interpolation of the distribution functions to each Lagrangian point is performed within this stencil. We use stencils of size $N_L\!=\!25$. Interpolation basis consists of cubic RBFs ($\phi(r)=r^3$) augmented with two-dimensional polynomial sets $\p_i$ of orders $i=2,4$:
	
	\begin{equation}\label{eq:poly_sets}
		\begin{aligned}
			&\p_2=\left\{1,x,y,x^2,xy,y^2\right\} \\
			&\p_4=\p_2 \cup \left\{x^3,x^2y,xy^2,y^3,x^4,x^3y,x^2y^2,xy^3,y^4\right\}.
		\end{aligned}
	\end{equation}
	The size of the polynomial subset of the interpolation basis introduced in Eq.~\eqref{eq:interpolant_value_alg} is equal to the power of $\p_i$: $N^P_L=\left|\p_i\right|$. Inverting local interpolation matrices $\bsym{\Phi}_L$ is performed using LU decomposition \cite{Crout1941,Banachiewicz1938}  with partial pivoting implemented in the \texttt{PartialPivLU} solver of the Eigen library \cite{Guennebaud2010}. The complete algorithm of the meshless LBM is presented in Appendix \ref{app:algorithm}.
	
	\section{Results}\label{sec:results}
	
	\setcounter{barmanEnum}{1}
	\addtocounter{barmanEnum}{1}
	
	We consider the Taylor-Green vortex flow as a benchmark test \cite{Taylor1937}. It is a popular flow problem devoid of walls and is thus suitable for the study of the convergence of numerical methods with no errors introduced by factors other than periodic boundary conditions. The domain is a two-dimensional square $D=[0,1]\times[0,1]$ with periodic boundary conditions. The velocity field of the flow is given by:
	\begin{equation}\label{eq:tgv_eqs}
		\begin{aligned}
			&u_{true}(t,x,y) = U_0\,\cos{2\pi x}\,\sin{2\pi y}\,e^{(-2\nu k^2t)} \\
			&v_{true}(t,x,y) = -U_0\,\sin{2\pi x}\,\cos{2\pi y}\,e^{(-2\nu k^2t)} \\
		\end{aligned},
		\quad
		t \ge 0
	\end{equation}
	with scaling parameter $k\!=\!2\pi L$ dependent on the domain side length $L\!=\!1$, the velocity magnitude $U_0\!=\!1$ and kinematic viscosity $\nu=1$.
	
\begin{figure}[h!]
		\centering
		\includegraphics[width=.8\linewidth]{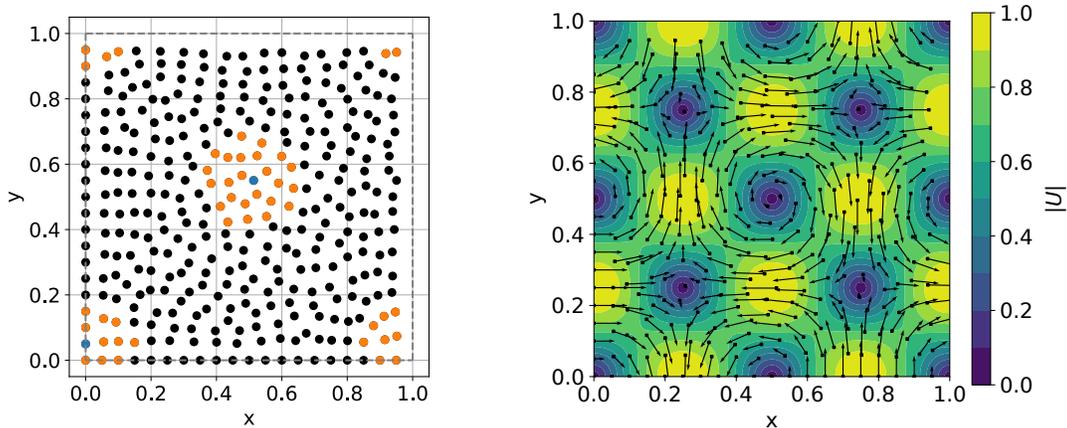}
		\caption{\textit{Left}: the $h\!=\!1/20$ meshless domain used in Taylor-Green vortices test. A dashedline denotes the periodic boundary. Orange points are the members of the stencils of the two blue points. \textit{Right}: a velocity map and vectors of the initial condition of the Taylor-Green vortex. For a clearer visualization of vectors a coarser discretization was used here compared to the left plot.}
		\label{fig:tgv_domain}
	\end{figure}

	In the meshless formulation, periodicity is implemented through a periodic search of the closest Eulerian neighbors and interpolation stencil members. The domain is discretized with an irregular point cloud obtained from the algorithm~\cite{Slak2019a} based on Poisson disk sampling implemented in the Medusa library~\cite{Slak2021}. After nodes generation, we run several relaxing iterations so that the nodes are distributed more uniformly, particularly to eliminate artifacts introduced by domain corners. Fig.~\ref{fig:tgv_domain} shows the point clouds used in the calculations.
	
\begin{figure}[h!]
		\centering
		\includegraphics[width=\linewidth]{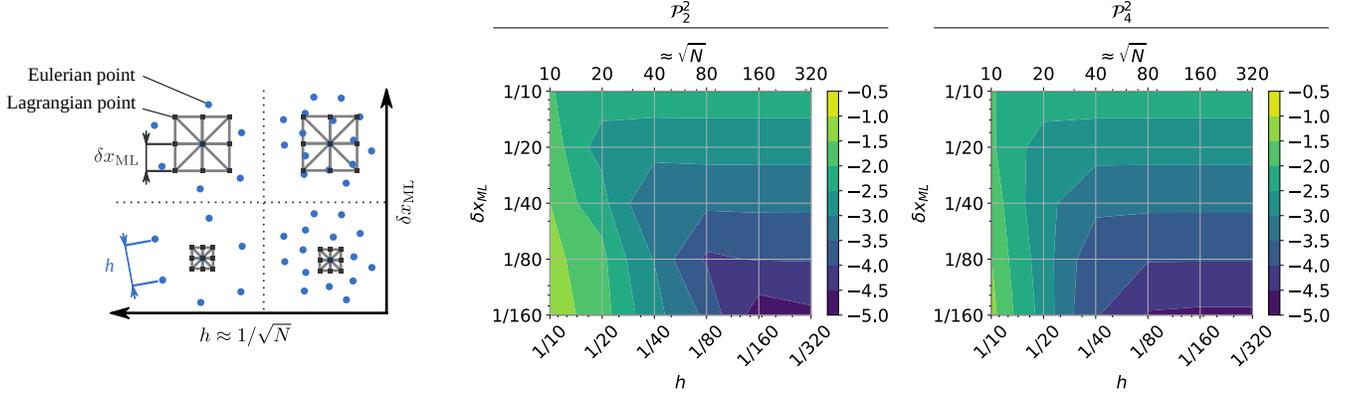} 	 
		\caption{\textit{Left}: in the meshless LBM the streaming distance $\delta x_\text{ML}$ (vertical axis) and the interpolation refinement $h$ (horizontal axis) can be altered independently. The choice of these two parameters influences errors introduced by the Boltzmann equation spatial discretization and the interpolation step, respectively. \textit{Middle and right}: $\text{log}_{10}$ of $L_2$ error norm (Eq.~\eqref{eq:L2norm}) of the $u$-component of velocity for the two orders of polynomial subset of the interpolation basis.}
		\label{fig:h_dx_explained}
	\end{figure}

	We use two measures of discretization refinement for the meshless LBM. The first is the square root of the number of Eulerian nodes in the domain, which is inversely proportional to the average distance between Eulerian nodes, $\sqrt{N} \sim h^{-1}$. This directly affects the interpolation accuracy. The other is the streaming distance $\delta x_\text{ML}$, which determines the LBM-related errors of the meshless formulation (see Fig.~\ref{fig:h_dx_explained}). In the standard LBM, the solution error depends only on the square root of the number of nodes in the domain $\sqrt{N}$. At the same time, $\sqrt{N}$ is a measure of computational and memory demands of each method and will be used as the main parameter for the comparison between them during the discussion on accuracy.
	
	For the meshless LBM we use $\delta x_{ML}=1/10,1/40,1/160$ and $h=1/10,1/20,1/40,1/80,1/160,1/320$. For the standard LBM we use $\delta x=1/10,1/20,1/40,1/80,1/160$. Such choice of $h$ and $\delta x$ gives a comparable number of nodes discretizing the domain in each method (the meshless discretizations used have approximately 15\% less nodes than the regular ones with $\delta x\!=\!h$). We also note that the values of $\delta x_\text{ML}$ form a subset of $\delta x$ values which allows for an easier investigation of the LBM-related errors of the meshless approach. To match time discretizations we set $\delta t=10^{-3}$ for $\delta x_\text{ML},\delta x=1/10$. To prevent the loss of the second order convergence in $\delta x_\text{ML},\delta x$ due to compressibility errors we perform diffusive scaling $\delta t \propto \delta x_\text{ML}^2,\delta x^2$. For both LBM variants we use $\tau=0.8$.
	
	\def\arraystretch{\bigstretch}
	
	The initial condition for the distribution functions was given by the equilibrium populations (Eq. \eqref{eq:feq}) parameterized with the macroscopic fields from Eq. \eqref{eq:tgv_eqs} at $t\!=\!0$ (see Fig.~\ref{fig:tgv_domain}).
	
	For the error measure we choose the normalized discrete norm of $u$-velocity error at time $t_{end}$ (see e.g. \cite{Nair2010,Lauritzen2012}):
	\begin{equation}\label{eq:L2norm}
		L_2 = \frac{ \sqrt{ \sum\limits_{i=1}^N\left(u_i^{t_{end}}-u^{t_{end}}_{i,true}\right)^2 } }
		{\sqrt{ \sum\limits_{i=1}^N \left(u^{t_{end}}_{i,true}\right)^2 } }
	\end{equation}
	where the subscript $i$ denotes the value of a variable at node $\boldsymbol{x}_i$. By transforming the exponential terms in Eq.~\eqref{eq:tgv_eqs} we choose:
	\begin{equation}
		t_{end} = \frac{\ln{10}}{2\nu k^2}
	\end{equation}
	which corresponds to the time, when the vortices have decayed to 10\% of their initial magnitude.
	
	The middle and the rightmost plots of Fig.~\ref{fig:h_dx_explained} show maps of $\log_{10}{L_2}$ for the meshless LBM solution in the space of the discretization parameters $\delta x_\text{ML}$, $\sqrt{N}$ and $h$. Each subplot concerns different order of the polynomial part of the interpolation basis. The vertical and bottom horizontal axes are aligned in the same manner as in the leftmost plot, and both maps share the same color scale. It is observed that increasing the order of the polynomial basis results in a decrease in the errors, which is most pronounced in the coarse $\sqrt{N}$ region. For a sufficiently small $\delta x_\text{ML}$ there is clear error convergence with $h$ refinement for both polynomial augmentations. However, for larger values of $\delta x_\text{ML}$, error convergence occurs only up to a certain point and then stops. A similar error behavior occurs for smaller values of $\sqrt{N}$ and refinement of $\delta x_\text{ML}$. For the most refined $\delta x_\text{ML}$ and largest $h$ even a divergence of the errors is observed, especially for $\p_2$. Tests performed with orders of the polynomial augmentation below 2 did not yield satisfactory results in the concerned range of $\delta x_\text{ML}$ and $\sqrt{N}$ values and thus will be excluded from further discussion.
	
\begin{figure}[h!]
		\centering
		\includegraphics[width=.6\linewidth]{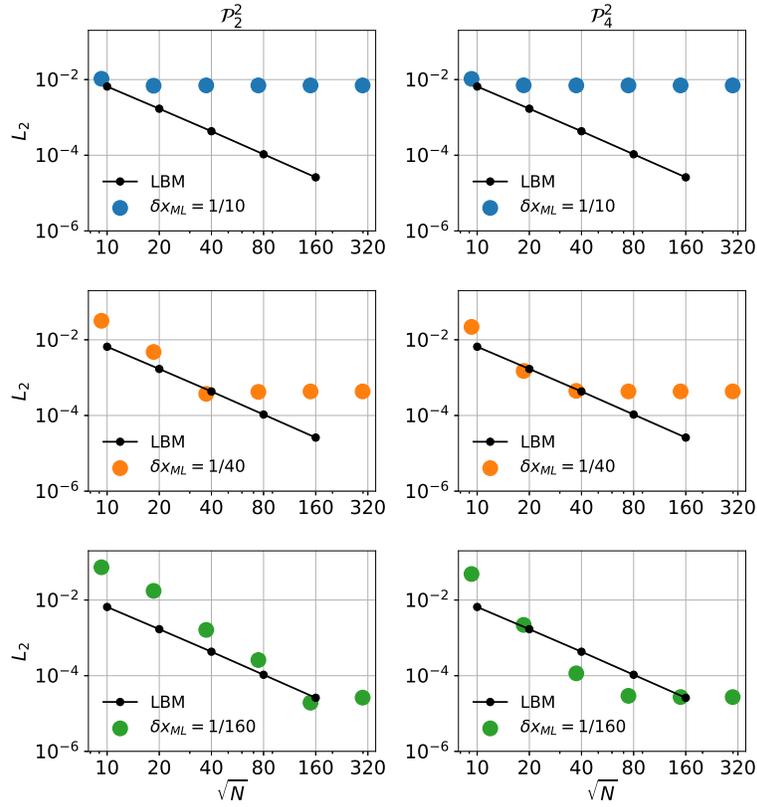}
		\caption{The convergence of $L_2$ error norm (Eq.~\eqref{eq:L2norm}) of the $u$-component of velocity for the Taylor-Green vortex test. Meshless LBM results ({\itshape large symbols}) are compared with standard LBM results ({\itshape small symbols}). A black solid line connecting the standard LBM results acts as a guide for the eye. Each column corresponds to a different polynomial subset $\p_i$ of the interpolation basis, each row corresponds to a different meshless LBM streaming distance $\delta x_\text{ML}$.}
		\label{fig:tgv_conv_P}
	\end{figure}

	Fig.~\ref{fig:tgv_conv_P} shows the same data compared with the error of the standard LBM (black symbols and a solid line). Each column corresponds to a different order of polynomial augmentation and each row concerns a separate $\delta x_\text{ML}$ value. Here we use $\sqrt{N}$ as the discretization refinement parameter for both methods - compare with the top horizontal axes in Fig.~\ref{fig:h_dx_explained}. We begin by analyzing $\delta x_\text{ML}\!=\!1/160$ results (bottom row). For $\p_4$ (right column) the cessation of meshless convergence is observed as early as for $\sqrt{N} \approx 80$. The meshless error reaches then the value of the standard LBM error with $\sqrt{N}\!=\!160$. This discretization refinement in the standard LBM gives the streaming distance $\delta x=1/160$ which is equal to the used meshless streaming distance $\delta x_\text{ML}=1/160$. From this point, continuing the refinement of $\sqrt{N}$ gives no further convergence of the meshless error. In addition, for $\p_4$ the meshless LBM achieves lower errors than the standard LBM for the same number of nodes (green symbols lying below the black curve) for $\sqrt{N} \! \approx \! 1/40$ and $1/80$. For $\p_2$ (left column), the meshless LBM error for $\delta x_\text{ML}\!=\!1/160$ reaches the mentioned LBM error only at $\sqrt{N} \! \approx \! 1/320$. However, it surpasses this value at $\sqrt{N} \approx 1/160$ (a green symbol lying lower than the bottommost black dot). This may be explained by the fact that interpolation and LBM-related errors of similar magnitudes cancel each other out. The meshless errors for $\delta x_\text{ML}=1/40$ (middle row) behave very similarly to those just described, and their lower bound in the case of a sufficiently refined interpolation discretization (large $\sqrt{N}$) is the standard LBM error at $\sqrt{N}=40$. In the case of $\delta x_\text{ML}=1/10$ (top row), the corresponding standard LBM error is reached already for the second coarsest $\sqrt{N}$. We also note that in the convergent regime of $\sqrt{N}$ the meshless results exhibit a higher rate of convergence than that of the standard LBM (which is of the 2nd order).
	
\begin{figure}[h!]
		\centering
		\includegraphics[width=.6\linewidth]{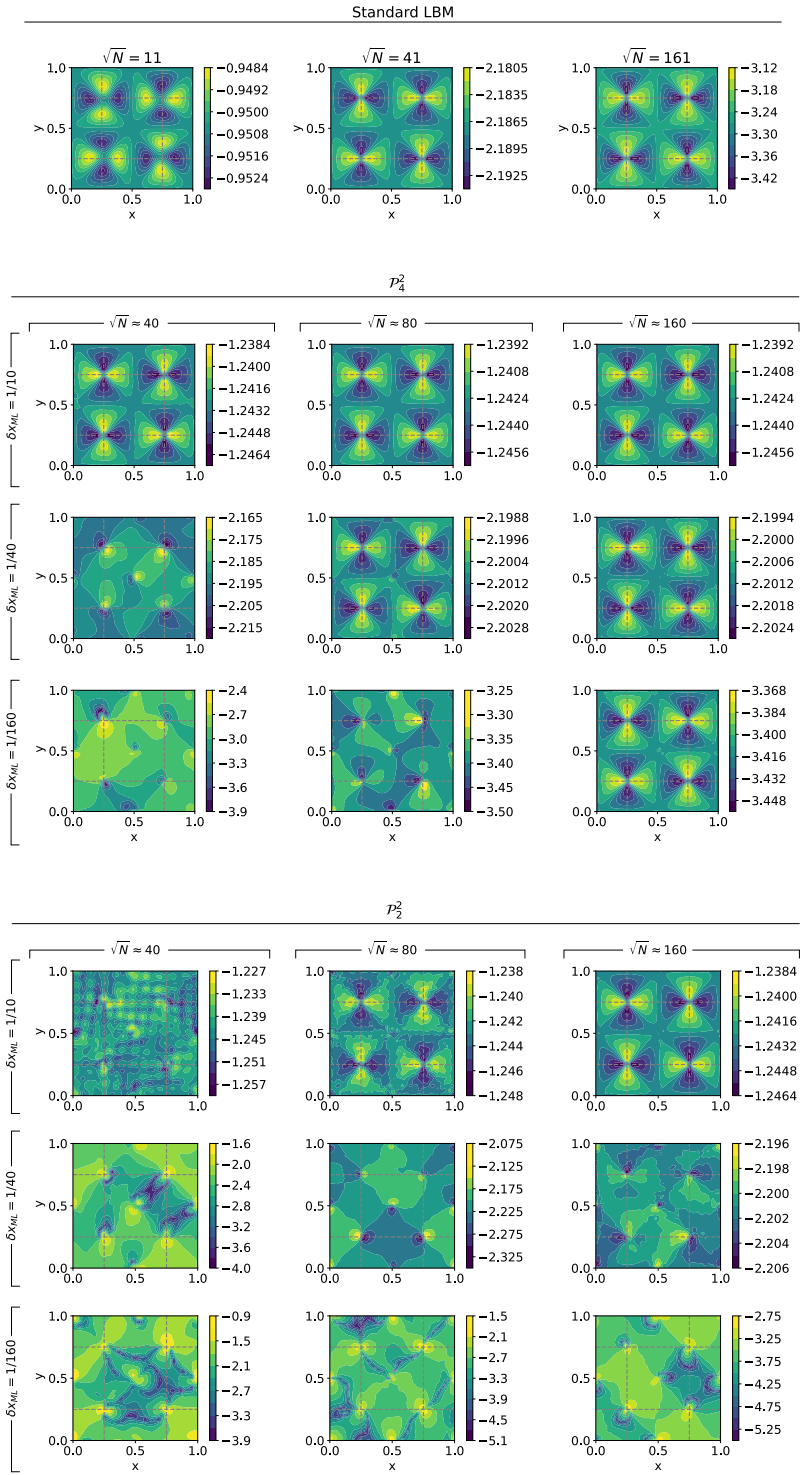}
		\caption{A spatial distribution of relative errors Eq.~\eqref{eq:vmag_error} of the velocity magnitude for the standard LBM (\textit{top}), and the meshless LBM with polynomial augmentation of the second (\textit{middle}) and fourth (\textit{bottom)} order.}
		\label{fig:spatial_error}
	\end{figure}

	Fig.~\ref{fig:spatial_error} shows $\log_{10}$ of the spatial distributions of the velocity magnitude relative error at time $t_\text{end}$:
	\begin{equation}\label{eq:vmag_error}
		e(\bsym{x}_i) \equiv e_i = \frac{\left|\>\left|U^{t_\text{end}}_i\right|-\left|U^{t_\text{end}}_{i,\text{true}}\right|\>\right|}{\left|U^{t_\text{end}}_{i,\text{true}}\right|}
		\> , \quad
		\left|U_i\right| = \sqrt{u^2_i+v^2_i}
	\end{equation}
	for the standard LBM and the meshless LBM obtained with the two polynomial augmentations $\p_2$ and $\p_4$. The standard LBM results use discretizations of $\sqrt{N}\!=\!11,41,161$ such that the nodes do not coincide with zero-velocity coordinates. For the meshless LBM, each row corresponds to a constant $\delta x_\text{ML}$ and each column corresponds to a constant $\sqrt{N}$. For the meshless LBM, 0.5\% of the points with the highest and lowest $e_i$ error values are excluded from the visualization. These are nodes with near-zero true velocity magnitudes. Standard LBM errors exhibit a spatial distribution that is symmetric with respect to a rotation about $\pi/2$ angle around the domain center $(0.5;0.5)$ regardless of $\sqrt{N}$. It can be explained by the same symmetry of the analytical solution, Eq.~\eqref{eq:tgv_eqs}. Four areas in the shape of $0.5 \times 0.5$ squares centered at stagnant velocity points (intersections of the dashed lines in Fig. ~\ref{fig:spatial_error}) can be clearly distinguished from the error patterns. Let us now relate this observation to the meshless LBM results, starting from the $\p_4$ case (the middle plot of Fig. ~\ref{fig:spatial_error}). The same spatial pattern of the $e_i$ error distribution is also observed here for the finest Eulerian points density $\sqrt{N} \! \approx \! 160$ (the rightmost column). With a decrease in $\sqrt{N}$ this spatial pattern seems to disappear in favor of some other spatial distribution of errors. This occurs at $\sqrt{N} \approx 40$ and $\sqrt{N} \approx 80$ for $\delta x_\text{ML}=1/160$. For $\p_2$ the error spatial pattern similar to that of the standard LBM is visible only for the finest $\sqrt{N}$ and the largest $\delta x_\text{ML}$. Regardless of the order of polynomial augmentation, whenever the standard LBM error pattern appears in the meshless solution, the magnitude of the errors is approximately the same as that in the standard LBM with a similar number of nodes $\sqrt{N}$. A possible explanation for these phenomena is that for relatively high accuracy of the interpolation (high order of $\p_i$ and fine $\sqrt{N}$) and large $\delta x_\text{ML}$, the LBM-related errors of the meshless solution are much higher than the interpolation errors, thus the LBM-like spatial pattern is visible. On the other hand, for low accuracy of the interpolation and low LBM-related errors (low order of $\p_i$, coarse $\sqrt{N}$, and small $\delta x_\text{ML}$) interpolation errors take over LBM-related errors and a pattern characteristic of the former is visible.
	
	To demonstrate the convergence of the discussed method in the presence of a body force and curved boundaries, we investigate a two-dimensional flow inside an annular channel. The channel walls are circles of radii $R_1$=1 and $R_2=2$ centered at the origin. The flow is forced with a constant acceleration $\bsym{g} = (0,g)$ in polar coordinates $(r,\phi)$. We implement the body force as an additional term in the collision operator, as suggested in~\cite{He1997a}:
	\begin{equation}
		f^\text{post}_k = f_k(t,\boldsymbol{x}) - \frac{1}{\tau}\left[f_k(t,\boldsymbol{x}) - f^\text{eq}_k(t,\boldsymbol{x})\right] + \omega_k \frac{\bsym{e}_k\cdot \bsym{g}}{c_s^2},
	\end{equation}
	and use the value of $g=10^{-6}$. We apply no-slip boundary condition to the stationary walls of the channel. We assume incompressibility of the fluid, zero radial component of velocity vectors in the whole domain, and rotational symmetry of the velocity and pressure field. The analytical solution of the Navier-Stokes equation with these assumptions is:
	\begin{equation}
		\begin{aligned}\label{eq:bch_solution}
			&u(r) = -G(r^2-\alpha r+ \frac{\beta}{r}) \\
			&\text{where} \\
			&G = \frac{g_\phi}{3\nu},\quad
			\alpha = \frac{R_1^2+R_1R_2+R_2^2}{R_1+R_2},\quad
			\beta = \frac{R_1^2R_2^2}{R_1+R_2},
		\end{aligned}
	\end{equation}
	We perform a series of meshless LBM simulations with the streaming distance $\delta x_\text{ML}\!=\!1/160$ and Eulerian nodes spacing $h=1/10,1/20,1/40,1/55,1/80$ and $1/110$. We set the relaxation time to $\tau=1$. We use a multireflection bounceback suggested by Ginzburg and {d'Humi{\`e}res}~\cite{Ginzburg2003}. In the meshless LBM, we place the boundary nodes exactly on the boundaries, so the multireflection bounceback reduces to assigning post-streaming population of the opposite lattice direction to the unknown populations:
	\begin{equation}
		f_k(t+1,\bsym{x}) = f_{k'}(t+1,\bsym{x}) = f^\text{post}_{k'}(t,\bsym{x}+\bsym{e}_k)
	\end{equation}
	The stencil size is $N_L=15$ and we use second-order polynomial augmentation for the interpolation. The initial condition is equilibrium distributions for zero macroscopic velocity at each node. The simulation was iterated until the relative $u$-velocity residual at a given timestep fell below $10^{-10}$ or up to $5\cdot10^5$ iterations. The measure of error is $L_2$ norm (Eq.~\eqref{eq:L2norm}) of the velocity magnitude.
	
\begin{figure}[h!]
		\centering
		\includegraphics[width=.49\linewidth]{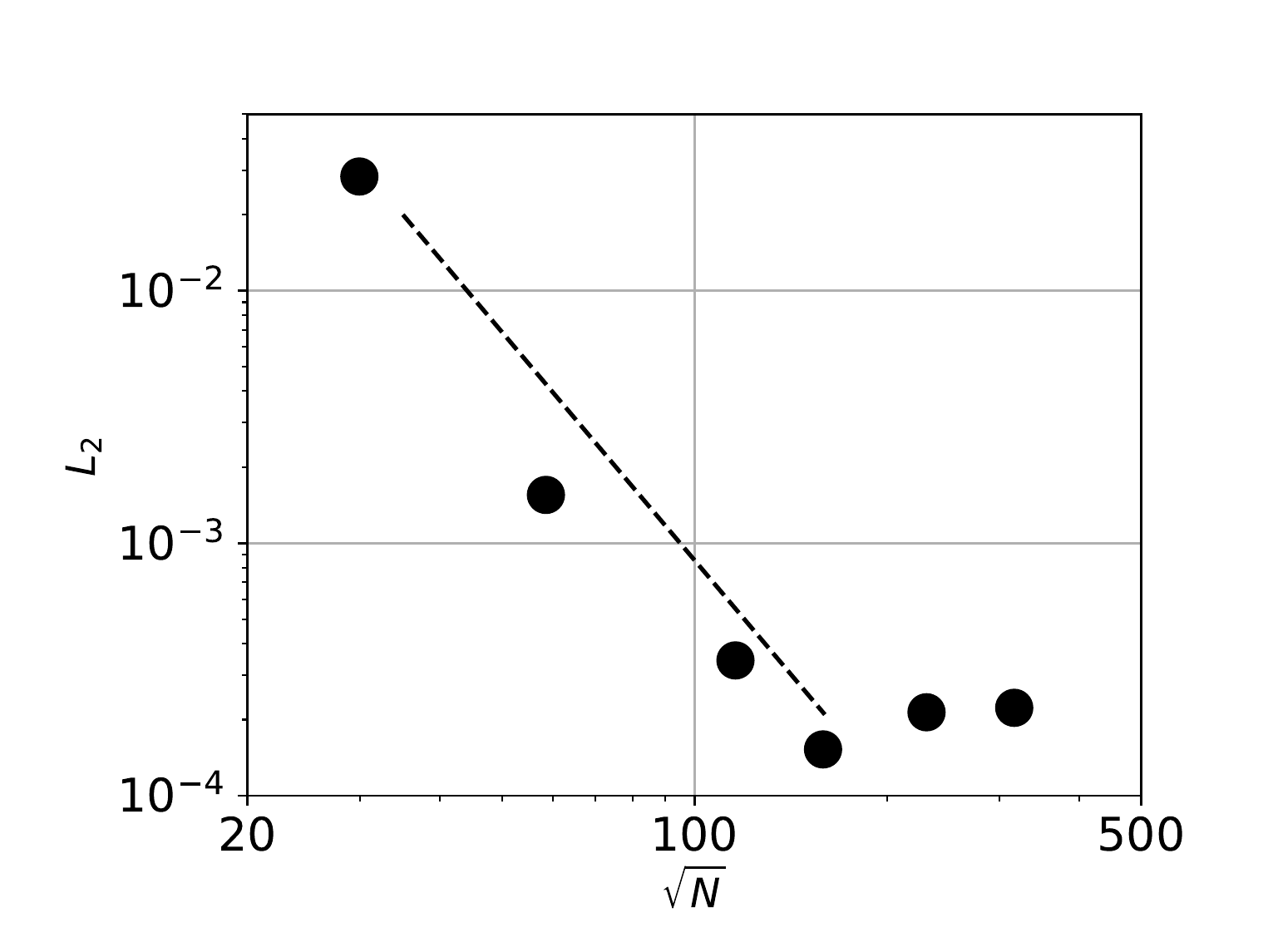}
		\includegraphics[width=.49\linewidth]{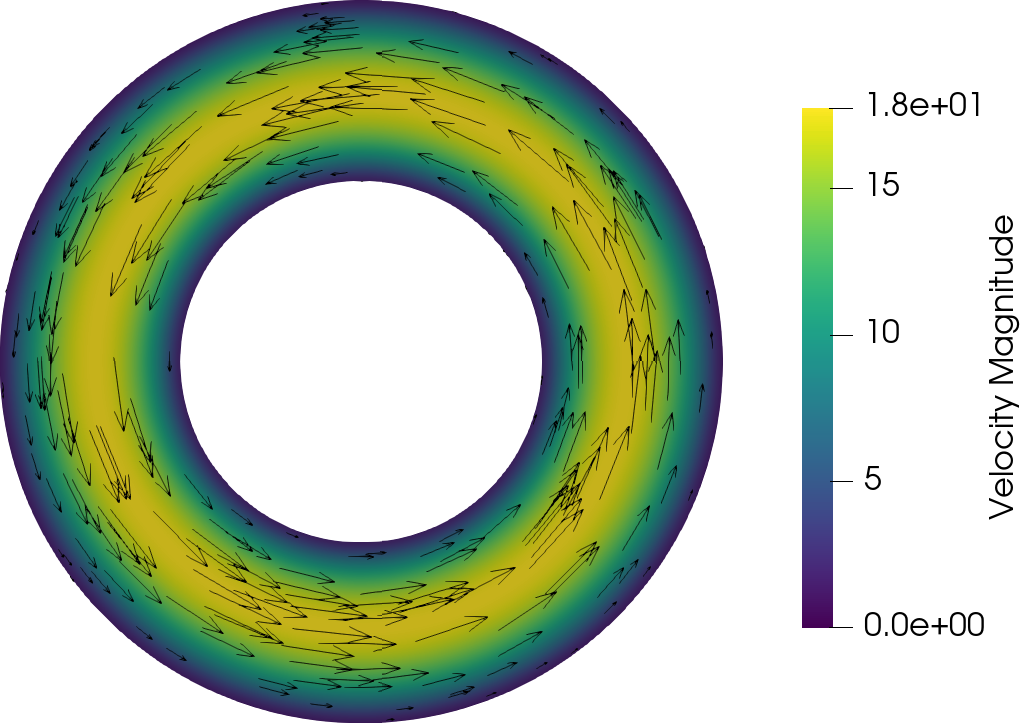}
		\caption{Results of the annular channel flow. \textit{Left}: the convergence of $L_2$ error norm (Eq.~\eqref{eq:L2norm}) of the $u$-component of velocity. Filled dots denote meshless LBM results and the dashed line represents the 3rd order of convergence. \textit{Right}: velocity magnitude field and a few chosen velocity vectors obtained from the meshless LBM simulation.}
		\label{fig:bch_conv}
	\end{figure}

	Fig.~\ref{fig:bch_conv} shows the convergence of ML-LBM error as a function of $\sqrt{N}$ (square root of the number of Eulerian nodes). Similarly to the Taylor-Green vortex test, the error convergence has an above-second-order rate for coarser grids, and for the finer ones it reaches its lower limit. One may stipulate that this lower limit would be the error of the standard LBM with $\delta x=1/160$ and third order (based on the convergence slope in Fig.~\ref{fig:bch_conv}) interpolation in the multireflexion bounceback.
	
	\section{Discussion}\label{sec:discussion}
	
	To explain the observations on the relation between the meshless and the standard LBM error in the absence of walls and body force, we take a look at Eq. (2.81) in \cite{Latt2007} where the standard LBM error is expressed as:
	\begin{equation}\label{eq:latt_err}
		E_{\text{LBM}}=\left(E_{\delta x}+E_{\delta t}+E_{Ma}\right) = \mathcal{O}(\delta x^2).
	\end{equation}
	Partial errors $E_{\delta x}$, $E_{\delta t}$, $E_{Ma}$ are related to space discretization, time discretization, and compressibility, respectively. The last equality above holds when the diffusive scaling of a timestep $\delta t \! \propto \! \delta x^2$ is used. Keeping in mind the semi-Lagrangian nature of the streaming step of the discussed meshless LBM one can introduce to Eq. \eqref{eq:latt_err} an error term characteristic to semi-Lagrangian numerical methods \cite{Liu2016,Falcone1998}:
	\begin{equation}\label{eq:semiLagrangian_err}
		E_{\text{SL}} = \mathcal{O}\left(\delta x^k + \frac{h^{p+1}}{\delta x}\right)
	\end{equation}
	\no where $k$ depends on the timestepping method used for solving the transport (streaming) step and $p$ is the order of interpolation. In such a way one arrives at the meshless LBM error dependent on the streaming distance $\delta x_\text{ML}$:
	\begin{equation}\label{eq:ML-LBM_err}
		E_{\text{RBF-LBM}}=\mathcal{O}\left(\delta x_\text{ML}^2 + \delta x_\text{ML}^k + \frac{h^{p+1}}{\delta x_\text{ML}}\right).
	\end{equation}
	From the above equation, several phenomena might be predicted. First, decreasing only the semi-Lagrangian part of the error ($\delta x_\text{ML}^k \! + \! h^{p+1}/\delta x_\text{ML}$), which in our case amounted to $\sqrt{N}$ and $\p_i$ refinement, leads to error convergence stagnation, since the LBM-related part ($\sim \! \delta x_\text{ML}^2$), ruled by $\delta x_\text{ML}$ and $\delta t$, does not fall then and eventually becomes much larger than the semi-Lagrangian streaming errors. Second, decreasing the streaming distance $\delta x_\text{ML}$ without accordingly decreasing the Eulerian points distance $h$ or increasing the interpolation order leads to the divergence of the last term in the asymptotic expression in Eq. \eqref{eq:ML-LBM_err}. Meshless LBM errors at a certain $\delta x_\text{ML}$ being lower than standard LBM errors with the same $\delta x=\delta x_\text{ML}$ may be explained by cancellations of errors introduced by the semi-Lagrangian and LBM terms in Eq.~\eqref{eq:ML-LBM_err}. This, however, is visible only when semi-Lagrangian and LBM errors are of similar orders of magnitude. Finally, the meshless LBM can be expected to give lower errors than the standard LBM with the same number of nodes $\sqrt{N}$ only when the order of interpolation $p+1$ is higher than 2 and with sufficiently small $\delta x_\text{ML}$.
	
	The detailed analysis of the error in the case when walls and a body force are present is beyond the scope of this work. Nevertheless, the presented results of the annular channel flow support the previous findings that the discussed meshless LBM variant is convergent in this case as well. The fact that the error convergence stops for some Eulerian discretization refinement suggests that the meshless LBM error relation from Eq.~\eqref{eq:ML-LBM_err} holds also in the presence of walls and body force. The exact value of the lower bound of the meshless LBM error should here depend not only on the standard LBM discretization $\delta x$, but also on the implementation of the body force and the no-slip boundaries. According to Guo and others \cite{Guo2002a} the forcing scheme used in the present study gives error in Navier-Stokes solution proportional to $\partial\bsym{g}/\partial t$, $\nabla \cdot \bsym{g}$ and $\nabla \cdot \bsym{gu}$. Since for the body force used in our study and for the resulting velocity field (Eq.~\eqref{eq:bch_solution}) the three terms are zero, we find it reasonable to assume that the contribution of the forcing scheme to the error of the macroscopic solution is negligible. Concerning the used bounceback scheme it is not trivial to assess the impact of the implemented multireflection method on the solution error. A comparison with the standard LBM solution with the third-order interpolation in the multireflection bounceback can give some hints into this matter. In practice, the implementation itself can be troublesome, since without the use of e.g. scattered nodes approximation methods, the interpolation stencils have to be adjusted to the local wall normal (even when the symmetries of the lattice discretizing the channel were exploited).
	
\begin{figure}[h!]
		\centering
		\includegraphics[width=.55\linewidth]{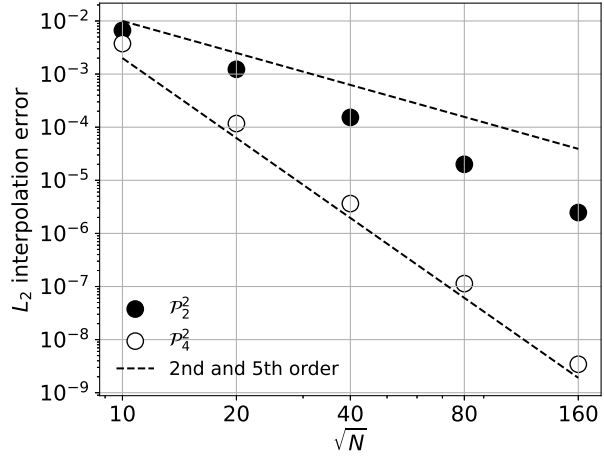}
		\caption{Convergence of the interpolation errors of a function $f(x,y) = 0.01 \sin{2\pi x}\cos{2\pi y}$ using the same point clouds, stencil size and polynomial subsets of the interpolation basis as in the meshless LBM setups. The interpolation takes place at Lagrangian nodes not coincident with the Eulerian points with the streaming distance $\delta x_\text{ML}=1/10$. Dashed lines denote the slopes of the 2nd and the 5th order convergence. The error measure is given by Eq. \eqref{eq:L2norm}.}
		\label{fig:interp_test}
	\end{figure}

	In \cite{He1996} the authors performed a convergence study of off-grid LBM with linear and quadratic interpolation on rectangular grids and noted the need for the interpolation to have at least the second order of convergence for interpolation errors do not outweigh LBM errors (compare the exponents $2$ and $p+1$ in Eq. \eqref{eq:ML-LBM_err}). From their results, one can estimate the orders of convergence as approximately equal to 1.6 when the quadratic interpolation is used. In the present study, the order of convergence for sole interpolation was higher than two for both polynomial augmentation sets $\p_2$, $\p_4$ (see Fig.~\ref{fig:interp_test}). As mentioned in Section~\ref{sec:results} using a lower-order polynomial augmentation resulted in the lack of a regular behavior of errors in the large $\sqrt{N}$ limit. However, a direct comparison of our results and \cite{He1996} is difficult since there the standard LBM results were used as a reference for calculating the off-grid LBM errors. Similarly, the authors of \cite{Kramer2017} used Lagrange polynomials on quadrilateral and hexahedral finite elements to perform the interpolation step in an off-grid LBM scheme. Our results are compliant with their findings on the above 2nd-order convergence of velocity error.
	
	The authors of \cite{Pribec2021} performed a similar error convergence study of an off-grid LBM using the Eulerian approach to solve the streaming step. They report the second-order convergence of $L_2$ velocity error norm in the Taylor-Green vortex test in the case of using point clouds of Cartesian arrangement (Fig.~3. therein). On the contrary to the present study, those results are obtained with the streaming distance scaled proportionally to the minimal distance between Eulerian points (i.e. $\delta t \propto h/|\bsym{c}_i|_\text{max}$ with constant $\bsym{c}_i$). Along with the lack of any analysis of the order of convergence of the used operator approximation techniques this makes it difficult to tell whether the second-order convergence of the mentioned $L_2$ norm comes from approximation or LBM-related errors convergence. On the other hand, when the authors of~\cite{Pribec2021} used irregular discretizations (Fig.~4 therein) they obtained above- or below-second-order convergence of the said error norm, depending on the approximation method. This may indicate that the observed convergence is related to the operator approximation error. The convergence results presented in two works of Musavi and Ashrafizaadeh (Fig.~1. in~\cite{Musavi2015} and Fig.~6. in~\cite{Musavi2016}) show an error saturation phenomenon similar to the one discussed in the present paper. Our and their results suggest that augmenting the standard LBM with some kind of numerical approximation leads to the saturation of errors regardless of the numerical scheme used to solve the streaming step. 
	We also note that our study concerns a wider range of Eulerian points spacings $h$ than in~\cite{Musavi2015,Musavi2016}, and thus we observe a cessation of convergence. We could expect the same in~\cite{Musavi2015,Musavi2016} if their study was continued for smaller $h$.
	However, one needs to be aware that Pribec~\cite{Pribec2021} and Musavi and Ashrafizaadeh~\cite{Musavi2015,Musavi2016} use Eulerian scheme for the streaming step and thus an application of the present conclusions to them should be done with caution.
	
	The further development of the MLBM methods is possible. They can be effectively used in complex LBM models e.g. Shan-Chen single-component multiphase model~\cite{Shan1993}. In such approaches the interactions between the fluid particles are encoded in the fully local collision operator, which is not affected by any changes made to the streaming step. Moreover, meshless formulation of LBM, with discretization nodes lying exactly on the boundaries, has a potential to simplify the calculation of wall normals, which is known to be problematic in the standard LBM, see e.g. work by Matyka et al.\cite{Matyka2013}.
	
	\section{Conclusions}\label{sec:conclusions}
	
	We show that in the absence of walls and body forces the error of the semi-Lagrangian meshless LBM presented in \cite{Lin2019} is convergent in the two discretization parameters: $\sqrt{N}$ which controls the interpolation error and $\delta x_\text{ML}$ accounting for the LBM-related error (Fig.~\ref{fig:h_dx_explained}). We show numerically that the lower bound on the meshless LBM error obtained with some value of $\delta x_\text{ML}$ is the value of the error of the standard LBM obtained with the discretization giving streaming distance $\delta x=\delta x_\text{ML}$ (Fig.~\ref{fig:tgv_conv_P}). We observe that increasing the order of interpolation via higher order polynomial augmentation results in a quicker convergence of the meshless LBM error to the said lower bound. In some cases, the meshless LBM can reach significantly lower errors than the standard LBM with the same number of nodes discretizing the domain. We show that the spatial distribution of meshless LBM errors reflects two regimes of the $\sqrt{N},\delta x_\text{ML}$ discretization: one dominated by interpolation errors and the other - by LBM-related errors (Fig.~\ref{fig:spatial_error}). Finally, we demonstrate that a similar behavior of the error norm convergence is visible when curved walls and a body force are present.
	
	\section*{Acknowledgements}\label{sec:acknowledgements}
	The authors would like to thank the members of Parallel and Distributed Systems Laboratory of Institute \textit{Jožef Stefan} in Ljubljana, Slovenia, for fruitful discussions and insightful remarks on the content of this work.
	
	Funded by National Science Centre, Poland under the OPUS call in the Weave programme 2021/43/I/ST3/00228.
	This research was funded in whole or in part by National Science Centre (2021/43/I/ST3/00228). For the purpose of Open Access,
	the author has applied a CC-BY public copyright licence to any Author Accepted Manuscript (AAM) version arising from this submission.
	
	
	\appendix
	\section{Meshless LBM algorithm}\label{app:algorithm}
	\begin{algorithm}[]
		\begin{algorithmic}[1]
			\State \textbf{Data:} Set of Eulerian nodes, simulation parameters;
			\For{all Eulerian points $\bsym{x}_i$}
			\State Initialize the macroscopic variables $\rho_i$, $u_i$ and $v_i$ (Eq.~\eqref{eq:tgv_eqs});
			\State Find stencil members $\bsym{x}_i^L$;
			\For{each lattice velocity $\bsym{c}_k$}
			\State Initialize the $k$-th distribution function with the equilibrium distribution $f_k^\text{eq}$ parameterized with the initial macroscopic variables (Eq.~\eqref{eq:feq});
			\State Determine the position of the $k$-th Lagrangian node $\bsym{x}_i+\delta t \bsym{c}_k$;
			\State Find the closest Eulerian neighbor of the $k$-th Lagrangian node;
			\State Calculate shape vector $\bsym{w}_L(\bsym{x}_i+\delta t \bsym{c}_k)$ for the $k$-th Lagrangian node (Eq.~\eqref{eq:shape_vector});
			\EndFor
			\EndFor
			\For{the prescribed number of timesteps}
			\Procedure{COLLIDE}{}
			\For{all Eulerian points $\bsym{x}_i$}
			\For{each lattice velocity $\bsym{c}_k$}
			\State Calculate the value of the equilibrium distribution $f^\text{eq}_k$ (Eq.~\eqref{eq:feq});
			\State Calculate the value of the post collision distribution $f^\text{post}_k$ (Eq.~\eqref{eq:LBM_collision});
			\EndFor
			\EndFor
			\EndProcedure
			\Procedure{STREAM}{}
			\For{all Eulerian points $\bsym{x}_i$}
			\For{each Lagrangian point $\bsym{x}_i+\delta t \bsym{c}_k$}
			\State Interpolate $f^\text{post}_{k'}$ to $\bsym{x}_i+\delta t \bsym{c}_k$ (Eq.~\eqref{eq:shape_vector});
			\State Overwrite $f_{k'}(\bsym{x}_i)$ (the $k'$-th distribution at $\bsym{x}_i$) with the interpolated $f^\text{post}_{k'}(\bsym{x}_i+\delta t \bsym{c}_k)$ (Eq.~\eqref{eq:LBM_streaming});
			\EndFor
			\State Update the values of the macroscopic variables (Eq.~\eqref{eq:macro_var});
			\EndFor
			\EndProcedure
			\EndFor
		\end{algorithmic}
		\caption{Semi-Lagrangian meshless LBM algorithm (in the absence of walls and body force)}
	\end{algorithm}
	\pagebreak

	\bibliography{../../../../library_zotero}

\begin{thebibliography}{10}
\urlstyle{rm}
\expandafter\ifx\csname url\endcsname\relax
  \def\url#1{\texttt{#1}}\fi
\expandafter\ifx\csname urlprefix\endcsname\relax\def\urlprefix{URL }\fi
\expandafter\ifx\csname doiprefix\endcsname\relax\def\doiprefix{DOI: }\fi
\providecommand{\bibinfo}[2]{#2}
\providecommand{\eprint}[2][]{\url{#2}}

\bibitem{Andrade1999}
\bibinfo{author}{Andrade, J.~S.}, \bibinfo{author}{Costa, U.~M.},
  \bibinfo{author}{Makse, H.~A.} \& \bibinfo{author}{Stanley, H.~E.}
\newblock \bibinfo{journal}{\bibinfo{title}{Role of {{Inertia}} on {{Fluid
  Flow}} through {{Disordered Porous Media}}}}.
\newblock {\emph{\JournalTitle{Physica A: Statistical Mechanics and its
  Applications}}} \textbf{\bibinfo{volume}{266}}, \bibinfo{pages}{420--424},
  \doiprefix\url{10.1016/S0378-4371(98)00624-4} (\bibinfo{year}{1999}).

\bibitem{Jin2016}
\bibinfo{author}{Jin, B.~J.}, \bibinfo{author}{Smith, A.~J.} \&
  \bibinfo{author}{Verkman, A.~S.}
\newblock \bibinfo{journal}{\bibinfo{title}{Spatial {{Model}} of {{Convective
  Solute Transport}} in {{Brain Extracellular Space Does Not Support}} a
  "{{Glymphatic}}" {{Mechanism}}}}.
\newblock {\emph{\JournalTitle{Journal of General Physiology}}}
  \textbf{\bibinfo{volume}{148}}, \bibinfo{pages}{489--501},
  \doiprefix\url{10.1085/jgp.201611684} (\bibinfo{year}{2016}).

\bibitem{Fu2018}
\bibinfo{author}{Fu, C.}, \bibinfo{author}{Uddin, M.} \&
  \bibinfo{author}{Robinson, A.~C.}
\newblock \bibinfo{journal}{\bibinfo{title}{Turbulence {{Modeling Effects}} on
  the {{CFD Predictions}} of {{Flow}} over a {{NASCAR Gen}} 6 {{Racecar}}}}.
\newblock {\emph{\JournalTitle{Journal of Wind Engineering and Industrial
  Aerodynamics}}} \textbf{\bibinfo{volume}{176}}, \bibinfo{pages}{98--111},
  \doiprefix\url{10.1016/j.jweia.2018.03.016} (\bibinfo{year}{2018}).

\bibitem{Jasak2019}
\bibinfo{author}{Jasak, H.}, \bibinfo{author}{Vuk{\v c}evi{\'c}, V.},
  \bibinfo{author}{Gatin, I.} \& \bibinfo{author}{Lalovi{\'c}, I.}
\newblock \bibinfo{journal}{\bibinfo{title}{{{CFD Validation}} and {{Grid
  Sensitivity Studies}} of {{Full Scale Ship Self Propulsion}}}}.
\newblock {\emph{\JournalTitle{International Journal of Naval Architecture and
  Ocean Engineering}}} \textbf{\bibinfo{volume}{11}}, \bibinfo{pages}{33--43},
  \doiprefix\url{10.1016/j.ijnaoe.2017.12.004} (\bibinfo{year}{2019}).

\bibitem{Baker2005}
\bibinfo{author}{Baker, T.~J.}
\newblock \bibinfo{journal}{\bibinfo{title}{Mesh {{Generation}}: {{Art}} or
  {{Science}}?}}
\newblock {\emph{\JournalTitle{Progress in Aerospace Sciences}}}
  \textbf{\bibinfo{volume}{41}}, \bibinfo{pages}{29--63},
  \doiprefix\url{10.1016/j.paerosci.2005.02.002} (\bibinfo{year}{2005}).

\bibitem{Liu2005}
\bibinfo{author}{Liu, G.~R.} \& \bibinfo{author}{Gu, Y.~T.}
\newblock \emph{\bibinfo{title}{An {{Introduction}} to {{Meshfree Methods}} and
  {{Their Programming}}}} (\bibinfo{publisher}{{Springer-Verlag}},
  \bibinfo{address}{{Berlin/Heidelberg}}, \bibinfo{year}{2005}).

\bibitem{Bayona2015}
\bibinfo{author}{Bayona, V.}, \bibinfo{author}{Flyer, N.},
  \bibinfo{author}{Lucas, G.~M.} \& \bibinfo{author}{Baumgaertner, A.~J.}
\newblock \bibinfo{journal}{\bibinfo{title}{A 3-{{D RBF-FD Solver}} for
  {{Modeling}} the {{Atmospheric Global Electric Circuit}} with {{Topography}}
  ({{GEC-RBFFD}} v1.0)}}.
\newblock {\emph{\JournalTitle{Geoscientific Model Development}}}
  \textbf{\bibinfo{volume}{8}}, \bibinfo{pages}{3007--3020},
  \doiprefix\url{10.5194/gmd-8-3007-2015} (\bibinfo{year}{2015}).

\bibitem{Slak2019}
\bibinfo{author}{Slak, J.} \& \bibinfo{author}{Kosec, G.}
\newblock \bibinfo{journal}{\bibinfo{title}{Adaptive {{Radial Basis
  Function}}\textendash{} {{Generated Finite Differences Method}} for {{Contact
  Problems}}}}.
\newblock {\emph{\JournalTitle{International Journal for Numerical Methods in
  Engineering}}} \textbf{\bibinfo{volume}{119}}, \bibinfo{pages}{661--686},
  \doiprefix\url{10.1002/nme.6067} (\bibinfo{year}{2019}).

\bibitem{Fornberg2015}
\bibinfo{author}{Fornberg, B.} \& \bibinfo{author}{Flyer, N.}
\newblock \bibinfo{journal}{\bibinfo{title}{Solving {{PDEs}} with {{Radial
  Basis Functions}}}}.
\newblock {\emph{\JournalTitle{Acta Numerica}}} \textbf{\bibinfo{volume}{24}},
  \bibinfo{pages}{215--258}, \doiprefix\url{10.1017/s0962492914000130}
  (\bibinfo{year}{2015}).

\bibitem{Succi2018}
\bibinfo{author}{Succi, S.}
\newblock \emph{\bibinfo{title}{The {{Lattice Boltzmann Equation}}: {{For
  Complex States}} of {{Flowing Matter}}}} (\bibinfo{publisher}{{Oxford
  University Press}}, \bibinfo{year}{2018}).

\bibitem{Matyka2021}
\bibinfo{author}{Matyka, M.} \& \bibinfo{author}{Dzikowski, M.}
\newblock \bibinfo{journal}{\bibinfo{title}{Memory-{{Efficient Lattice
  Boltzmann Method}} for {{Low Reynolds Number Flows}}}}.
\newblock {\emph{\JournalTitle{Computer Physics Communications}}}
  \textbf{\bibinfo{volume}{267}}, \bibinfo{pages}{108044},
  \doiprefix\url{10.1016/j.cpc.2021.108044} (\bibinfo{year}{2021}).

\bibitem{Koponen1997}
\bibinfo{author}{Koponen, A.}, \bibinfo{author}{Kataja, M.} \&
  \bibinfo{author}{Timonen, J.}
\newblock \bibinfo{journal}{\bibinfo{title}{Permeability and {{Effective
  Porosity}} of {{Porous Media}}}}.
\newblock {\emph{\JournalTitle{Physical Review E - Statistical Physics,
  Plasmas, Fluids, and Related Interdisciplinary Topics}}}
  \textbf{\bibinfo{volume}{56}}, \bibinfo{pages}{3319--3325},
  \doiprefix\url{10.1103/PhysRevE.56.3319} (\bibinfo{year}{1997}).

\bibitem{Paradis2013}
\bibinfo{author}{Paradis, H.}, \bibinfo{author}{Grigoropoulos, C.} \&
  \bibinfo{author}{Sunden, B.}
\newblock \bibinfo{title}{Lattice {{Boltzmann Modeling}} for {{Analysis}} of
  {{Water-Splitting Over Nanorods With Emphasis}} on {{Reactive Mass
  Transport}}}.
\newblock In \emph{\bibinfo{booktitle}{{{ASME}} 2013 11th {{International
  Conference}} on {{Nanochannels}}, {{Microchannels}} and {{Minichannels}},
  {{ICNMM}} 2013}}, \doiprefix\url{10.1115/ICNMM2013-73098}
  (\bibinfo{year}{2013}).

\bibitem{Gu2021}
\bibinfo{author}{Gu, Q.}, \bibinfo{author}{Liu, H.} \& \bibinfo{author}{Wu, L.}
\newblock \bibinfo{journal}{\bibinfo{title}{Preferential imbibition in a
  dual-permeability pore network}}.
\newblock {\emph{\JournalTitle{Journal of Fluid Mechanics}}}
  \textbf{\bibinfo{volume}{915}}, \bibinfo{pages}{A138},
  \doiprefix\url{10.1017/jfm.2021.174} (\bibinfo{year}{2021}).

\bibitem{Falcucci2013}
\bibinfo{author}{Falcucci, G.}, \bibinfo{author}{Jannelli, E.},
  \bibinfo{author}{Ubertini, S.} \& \bibinfo{author}{Succi, S.}
\newblock \bibinfo{journal}{\bibinfo{title}{Direct numerical evidence of
  stress-induced cavitation}}.
\newblock {\emph{\JournalTitle{Journal of Fluid Mechanics}}}
  \textbf{\bibinfo{volume}{728}}, \bibinfo{pages}{362--375},
  \doiprefix\url{10.1017/jfm.2013.271} (\bibinfo{year}{2013}).

\bibitem{Coelho2018}
\bibinfo{author}{Coelho, R.~C.} \& \bibinfo{author}{Doria, M.~M.}
\newblock \bibinfo{journal}{\bibinfo{title}{Lattice {{Boltzmann Method}} for
  {{Semiclassical Fluids}}}}.
\newblock {\emph{\JournalTitle{Computers and Fluids}}}
  \textbf{\bibinfo{volume}{165}}, \bibinfo{pages}{144--159},
  \doiprefix\url{10.1016/j.compfluid.2018.01.019} (\bibinfo{year}{2018}).

\bibitem{Mendoza2010}
\bibinfo{author}{Mendoza, M.}, \bibinfo{author}{Boghosian, B.~M.},
  \bibinfo{author}{Herrmann, H.~J.} \& \bibinfo{author}{Succi, S.}
\newblock \bibinfo{journal}{\bibinfo{title}{Fast {{Lattice Boltzmann Solver}}
  for {{Relativistic Hydrodynamics}}}}.
\newblock {\emph{\JournalTitle{Physical Review Letters}}}
  \textbf{\bibinfo{volume}{105}},
  \doiprefix\url{10.1103/PhysRevLett.105.014502} (\bibinfo{year}{2010}).

\bibitem{He1997}
\bibinfo{author}{He, X.} \& \bibinfo{author}{Doolen, G.}
\newblock \bibinfo{journal}{\bibinfo{title}{Lattice {{Boltzmann Method}} on
  {{Curvilinear Coordinates System}}: {{Flow}} around a {{Circular
  Cylinder}}}}.
\newblock {\emph{\JournalTitle{Journal of Computational Physics}}}
  \textbf{\bibinfo{volume}{134}}, \bibinfo{pages}{306--315},
  \doiprefix\url{10.1006/jcph.1997.5709} (\bibinfo{year}{1997}).

\bibitem{He1996}
\bibinfo{author}{He, X.}, \bibinfo{author}{Luo, L.~S.} \&
  \bibinfo{author}{Dembo, M.}
\newblock \bibinfo{journal}{\bibinfo{title}{Some {{Progress}} in {{Lattice
  Boltzmann Method}}. {{Part I}}. {{Nonuniform Mesh Grids}}}}.
\newblock {\emph{\JournalTitle{Journal of Computational Physics}}}
  \textbf{\bibinfo{volume}{129}}, \bibinfo{pages}{357--363},
  \doiprefix\url{10.1006/jcph.1996.0255} (\bibinfo{year}{1996}).

\bibitem{Kramer2017}
\bibinfo{author}{Kr{\"a}mer, A.}, \bibinfo{author}{K{\"u}llmer, K.},
  \bibinfo{author}{Reith, D.}, \bibinfo{author}{Joppich, W.} \&
  \bibinfo{author}{Foysi, H.}
\newblock \bibinfo{journal}{\bibinfo{title}{Semi-{{Lagrangian}} off-{{Lattice
  Boltzmann Method}} for {{Weakly Compressible Flows}}}}.
\newblock {\emph{\JournalTitle{Physical Review E}}}
  \textbf{\bibinfo{volume}{95}}, \bibinfo{pages}{1--12},
  \doiprefix\url{10.1103/PhysRevE.95.023305} (\bibinfo{year}{2017}).

\bibitem{Misztal2015}
\bibinfo{author}{Misztal, M.~K.} \emph{et~al.}
\newblock \bibinfo{journal}{\bibinfo{title}{Simulating {{Anomalous Dispersion}}
  in {{Porous Media Using}} the {{Unstructured Lattice Boltzmann Method}}}}.
\newblock {\emph{\JournalTitle{Frontiers in Physics}}}
  \textbf{\bibinfo{volume}{3}}, \bibinfo{pages}{1--9},
  \doiprefix\url{10.3389/fphy.2015.00050} (\bibinfo{year}{2015}).

\bibitem{Bardow2006}
\bibinfo{author}{Bardow, A.}, \bibinfo{author}{Karlin, I.~V.} \&
  \bibinfo{author}{Gusev, A.~A.}
\newblock \bibinfo{journal}{\bibinfo{title}{General {{Characteristic-Based
  Algorithm}} for off-{{Lattice Boltzmann Simulations}}}}.
\newblock {\emph{\JournalTitle{Europhysics Letters}}}
  \textbf{\bibinfo{volume}{75}}, \bibinfo{pages}{434--440},
  \doiprefix\url{10.1209/epl/i2006-10138-1} (\bibinfo{year}{2006}).

\bibitem{Lee2003}
\bibinfo{author}{Lee, T.} \& \bibinfo{author}{Lin, C.~L.}
\newblock \bibinfo{journal}{\bibinfo{title}{An {{Eulerian Description}} of the
  {{Streaming Process}} in the {{Lattice Boltzmann Equation}}}}.
\newblock {\emph{\JournalTitle{Journal of Computational Physics}}}
  \textbf{\bibinfo{volume}{185}}, \bibinfo{pages}{445--471},
  \doiprefix\url{10.1016/S0021-9991(02)00065-7} (\bibinfo{year}{2003}).

\bibitem{Lin2019}
\bibinfo{author}{Lin, X.}, \bibinfo{author}{Wu, J.} \& \bibinfo{author}{Zhang,
  T.}
\newblock \bibinfo{journal}{\bibinfo{title}{A {{Mesh-Free Radial Basis
  Function}}\textendash{} {{Based Semi-Lagrangian Lattice Boltzmann Method}}
  for {{Incompressible Flows}}}}.
\newblock {\emph{\JournalTitle{International Journal for Numerical Methods in
  Fluids}}} \textbf{\bibinfo{volume}{91}}, \bibinfo{pages}{198--211},
  \doiprefix\url{10.1002/fld.4749} (\bibinfo{year}{2019}).

\bibitem{Slak2019a}
\bibinfo{author}{Slak, J.} \& \bibinfo{author}{Kosec, G.}
\newblock \bibinfo{journal}{\bibinfo{title}{On {{Generation}} of {{Node
  Distributions}} for {{Meshless PDE Discretizations}}}}.
\newblock {\emph{\JournalTitle{SIAM Journal on Scientific Computing}}}
  \textbf{\bibinfo{volume}{41}}, \bibinfo{pages}{A3202--A3229},
  \doiprefix\url{10.1137/18M1231456} (\bibinfo{year}{2019}).

\bibitem{Kruger2017}
\bibinfo{author}{Kr{\"u}ger, T.} \emph{et~al.}
\newblock \emph{\bibinfo{title}{The {{Lattice Boltzmann Method}}:
  {{Principles}} and {{Practice}}}}.
\newblock Graduate {{Texts}} in {{Physics}} (\bibinfo{publisher}{{Springer
  International Publishing}}, \bibinfo{address}{{Cham}}, \bibinfo{year}{2017}).

\bibitem{Bhatnagar1954}
\bibinfo{author}{Bhatnagar, P.~L.}, \bibinfo{author}{Gross, E.~P.} \&
  \bibinfo{author}{Krook, M.}
\newblock \bibinfo{journal}{\bibinfo{title}{A {{Model}} for {{Collision
  Processes}} in {{Gases}}. {{I}}. {{Small Amplitude Processes}} in {{Charged}}
  and {{Neutral One-Component Systems}}}}.
\newblock {\emph{\JournalTitle{Phys. Rev.}}} \textbf{\bibinfo{volume}{94}},
  \bibinfo{pages}{511--525}, \doiprefix\url{10.1103/PhysRev.94.511}
  (\bibinfo{year}{1954}).

\bibitem{Crout1941}
\bibinfo{author}{Crout, P.~D.}
\newblock \bibinfo{journal}{\bibinfo{title}{A {{Short Method}} for {{Evaluating
  Determinants}} and {{Solving Systems}} of {{Linear Equations}} with {{Real}}
  or {{Complex Coefficients}}}}.
\newblock {\emph{\JournalTitle{Transactions of the American Institute of
  Electrical Engineers}}} \textbf{\bibinfo{volume}{60}},
  \bibinfo{pages}{1235--1240} (\bibinfo{year}{1941}).

\bibitem{Banachiewicz1938}
\bibinfo{author}{Banachiewicz, T.}
\newblock \bibinfo{journal}{\bibinfo{title}{M\'ethode de {{R\'esolution
  Num\'erique Des \'Equations Lin\'eaires}}, {{Du Calcul Des D\'eterminants}}
  et {{Des Inverses}}, et de {{R\'eduction Des Formes Quadratique}}}}.
\newblock {\emph{\JournalTitle{Bull. Acad. Pol. Ser. A.}}}
  \bibinfo{pages}{393--404} (\bibinfo{year}{1938}).

\bibitem{Guennebaud2010}
\bibinfo{author}{Guennebaud, G.}, \bibinfo{author}{Jacob, B.} \emph{et~al.}
\newblock \bibinfo{title}{Eigen {{V3}}} (\bibinfo{year}{2010}).

\bibitem{Taylor1937}
\bibinfo{author}{Taylor, G.~I.} \& \bibinfo{author}{Green, A.~E.}
\newblock \bibinfo{journal}{\bibinfo{title}{Mechanism of the {{Production}} of
  {{Small Eddies}} from {{Large Ones}}}}.
\newblock {\emph{\JournalTitle{Proceedings of the Royal Society of London.
  Series A - Mathematical and Physical Sciences}}}
  \textbf{\bibinfo{volume}{158}}, \bibinfo{pages}{499--521},
  \doiprefix\url{10.1098/rspa.1937.0036} (\bibinfo{year}{1937}).

\bibitem{Slak2021}
\bibinfo{author}{Slak, J.} \& \bibinfo{author}{Kosec, G.}
\newblock \bibinfo{journal}{\bibinfo{title}{Medusa: {{A C}}++ {{Library}} for
  {{Solving PDEs Using Strong Form Mesh-free Methods}}}}.
\newblock {\emph{\JournalTitle{ACM Transactions on Mathematical Software}}}
  \textbf{\bibinfo{volume}{47}}, \doiprefix\url{10.1145/3450966}
  (\bibinfo{year}{2021}).

\bibitem{Nair2010}
\bibinfo{author}{Nair, R.~D.} \& \bibinfo{author}{Lauritzen, P.~H.}
\newblock \bibinfo{journal}{\bibinfo{title}{A {{Class}} of {{Deformational Flow
  Test Cases}} for {{Linear Transport Problems}} on the {{Sphere}}}}.
\newblock {\emph{\JournalTitle{Journal of Computational Physics}}}
  \textbf{\bibinfo{volume}{229}}, \bibinfo{pages}{8868--8887},
  \doiprefix\url{10.1016/j.jcp.2010.08.014} (\bibinfo{year}{2010}).

\bibitem{Lauritzen2012}
\bibinfo{author}{Lauritzen, P.~H.}, \bibinfo{author}{Skamarock, W.~C.},
  \bibinfo{author}{Prather, M.~J.} \& \bibinfo{author}{Taylor, M.~A.}
\newblock \bibinfo{journal}{\bibinfo{title}{A {{Standard Test Case Suite}} for
  {{Two-Dimensional Linear Transport}} on the {{Sphere}}}}.
\newblock {\emph{\JournalTitle{Geoscientific Model Development}}}
  \textbf{\bibinfo{volume}{5}}, \bibinfo{pages}{887--901},
  \doiprefix\url{10.5194/gmd-5-887-2012} (\bibinfo{year}{2012}).

\bibitem{He1997a}
\bibinfo{author}{He, X.}, \bibinfo{author}{Zou, Q.}, \bibinfo{author}{Luo,
  L.-S.} \& \bibinfo{author}{Dembo, M.}
\newblock \bibinfo{journal}{\bibinfo{title}{Analytic solutions of simple flows
  and analysis of nonslip boundary conditions for the lattice {{Boltzmann BGK}}
  model}}.
\newblock {\emph{\JournalTitle{Journal of Statistical Physics}}}
  \textbf{\bibinfo{volume}{87}}, \bibinfo{pages}{115--136},
  \doiprefix\url{10.1007/BF02181482} (\bibinfo{year}{1997}).

\bibitem{Ginzburg2003}
\bibinfo{author}{Ginzburg, I.} \& \bibinfo{author}{{d'Humi{\`e}res}, D.}
\newblock \bibinfo{journal}{\bibinfo{title}{Multireflection {{Boundary
  Conditions}} for {{Lattice Boltzmann Models}}}}.
\newblock {\emph{\JournalTitle{Physical Review E}}}
  \textbf{\bibinfo{volume}{68}}, \bibinfo{pages}{066614},
  \doiprefix\url{10.1103/PhysRevE.68.066614} (\bibinfo{year}{2003}).

\bibitem{Latt2007}
\bibinfo{author}{Latt, J.}
\newblock \emph{\bibinfo{title}{Hydrodynamic {{Limit}} of {{Lattice Boltzmann
  Equations}}}}.
\newblock Ph.D. thesis (\bibinfo{year}{2007}).
\newblock \doiprefix\url{10.13097/archive-ouverte/unige:464}.

\bibitem{Liu2016}
\bibinfo{author}{Liu, L.} \& \bibinfo{author}{Becerra, M.}
\newblock \bibinfo{journal}{\bibinfo{title}{An {{Efficient Semi-Lagrangian
  Algorithm}} for {{Simulation}} of {{Corona Discharges}}: {{The Position-State
  Separation Method}}}}.
\newblock {\emph{\JournalTitle{IEEE Transactions on Plasma Science}}}
  \textbf{\bibinfo{volume}{44}}, \bibinfo{pages}{2822--2831},
  \doiprefix\url{10.1109/TPS.2016.2609504} (\bibinfo{year}{2016}).

\bibitem{Falcone1998}
\bibinfo{author}{Falcone, M.} \& \bibinfo{author}{Ferretti, R.}
\newblock \emph{\bibinfo{title}{Convergence {{Analysis}} for a {{Class}} of
  {{High-Order Semi-Lagrangian Advection Schemes}}}}, vol.~\bibinfo{volume}{35}
  (\bibinfo{publisher}{{SIAM Journal on Numerical Analysis}},
  \bibinfo{year}{1998}).

\bibitem{Guo2002a}
\bibinfo{author}{Guo, Z.}, \bibinfo{author}{Zheng, C.} \& \bibinfo{author}{Shi,
  B.}
\newblock \bibinfo{journal}{\bibinfo{title}{Discrete lattice effects on the
  forcing term in the lattice {{Boltzmann}} method}}.
\newblock {\emph{\JournalTitle{Physical Review E}}}
  \textbf{\bibinfo{volume}{65}}, \bibinfo{pages}{046308},
  \doiprefix\url{10.1103/PhysRevE.65.046308} (\bibinfo{year}{2002}).

\bibitem{Pribec2021}
\bibinfo{author}{Pribec, I.}, \bibinfo{author}{Becker, T.} \&
  \bibinfo{author}{Fattahi, E.}
\newblock \bibinfo{journal}{\bibinfo{title}{A {{Strong-Form Off-Lattice
  Boltzmann Method}} for {{Irregular Point Clouds}}}}.
\newblock {\emph{\JournalTitle{Symmetry}}} \textbf{\bibinfo{volume}{13}},
  \bibinfo{pages}{1802}, \doiprefix\url{10.3390/sym13101802}
  (\bibinfo{year}{2021}).

\bibitem{Musavi2015}
\bibinfo{author}{Musavi, S.~H.} \& \bibinfo{author}{Ashrafizaadeh, M.}
\newblock \bibinfo{journal}{\bibinfo{title}{Meshless {{Lattice Boltzmann
  Method}} for the {{Simulation}} of {{Fluid Flows}}}}.
\newblock {\emph{\JournalTitle{Physical Review E - Statistical, Nonlinear, and
  Soft Matter Physics}}} \textbf{\bibinfo{volume}{91}},
  \doiprefix\url{10.1103/PhysRevE.91.023310} (\bibinfo{year}{2015}).

\bibitem{Musavi2016}
\bibinfo{author}{Musavi, S.~H.} \& \bibinfo{author}{Ashrafizaadeh, M.}
\newblock \bibinfo{journal}{\bibinfo{title}{A mesh-free lattice {{Boltzmann}}
  solver for flows in complex geometries}}.
\newblock {\emph{\JournalTitle{International Journal of Heat and Fluid Flow}}}
  \textbf{\bibinfo{volume}{59}}, \bibinfo{pages}{10--19},
  \doiprefix\url{10.1016/j.ijheatfluidflow.2016.01.006} (\bibinfo{year}{2016}).

\bibitem{Shan1993}
\bibinfo{author}{Shan, X.} \& \bibinfo{author}{Chen, H.}
\newblock \bibinfo{journal}{\bibinfo{title}{Lattice {{Boltzmann Model}} for
  {{Simulating Flows}} with {{Multi Phases}} and components}}.
\newblock {\emph{\JournalTitle{Physical Review E}}}
  \textbf{\bibinfo{volume}{47}}, \bibinfo{pages}{1815--1819}
  (\bibinfo{year}{1993}).

\bibitem{Matyka2013}
\bibinfo{author}{Matyka, M.}, \bibinfo{author}{Koza, Z.} \&
  \bibinfo{author}{Miros{\l}aw, {\L}.}
\newblock \bibinfo{journal}{\bibinfo{title}{Wall orientation and shear stress
  in the lattice {{Boltzmann}} model}}.
\newblock {\emph{\JournalTitle{Computers \& Fluids}}}
  \textbf{\bibinfo{volume}{73}}, \bibinfo{pages}{115--123},
  \doiprefix\url{10.1016/j.compfluid.2012.12.018} (\bibinfo{year}{2013}).

\end{thebibliography}
	
	\pagebreak
	
	\renewcommand\thefigure{\arabic{figure}}
	
	\onecolumn		
	
\end{document}